\documentclass{optica-article}

\journal{opticajournal} 

\hypersetup{
    colorlinks=true,       
    linkcolor=red,          
   citecolor=magenta,        
    filecolor=magenta,      
    urlcolor=cyan,           
    runcolor=cyan
}

\articletype{Research Article}

\usepackage{subcaption}
\usepackage{lineno}
\usepackage{siunitx}

\begin{document}

\title{Picosecond synchronization of mode-locked lasers for metropolitan-scale quantum networks}

\author{Cory Nunn,\authormark{1,*} Nijil Lal,\authormark{1}, Ivan Burenkov,\authormark{1,2} Ya-Shian Li-Baboud,\authormark{1} Paulina S. Kuo,\authormark{1} Thomas Gerrits,\authormark{1} and Sergey V. Polyakov\authormark{1,3}}

\address{\authormark{1}National Institute of Standards and Technology, Gaithersburg, MD 20899, USA\\
\authormark{2}Joint Quantum Institute \& University of Maryland, College Park MD 20742, USA\\
\authormark{3}Department of Physics, University of Maryland, College Park, MD 20742, USA}

\email{\authormark{*}cory.nunn@nist.gov}

\begin{abstract*} 
We demonstrate picosecond-level synchronization of two actively mode-locked Ti:Sapphire lasers via the White Rabbit Precision Time Protocol (WR-PTP), tested over 120~km of deployed optical fiber. This synchronization capability, in combination with the highly single-mode, transform-limited pulses produced by each laser, enables their use as pump lasers for indistinguishable photon sources at remote locations in a quantum network. Here, the WR-PTP serves as a scalable network synchronization protocol, and its performance is compared to traditional methods of local synchronization.
We demonstrate pulse-to-pulse synchronization better than 3~ps and time deviation (TDEV) values below 4~ps for all averaging times up to 10~s. With a designed coherence time of 35~ps for single photon sources utilizing these lasers, the achievable temporal overlap corresponds to 98~\%  Hong-Ou-Mandel (HOM) interference visibility between independent sources.

\end{abstract*}

\section{Introduction}\label{sec:intro}

Synchronization is fundamentally important for a range of quantum network applications, including quantum key distribution~\cite{bienfang2004qkd_sync}, entanglement distribution~\cite{cirac1997entdist,kim2022timebin}, and entanglement swapping~\cite{jin2015swapping,sun2017swapping,pant2019swapping,wang2022entdist}.

In particular, protocols like swapping rely on the interference of single photons from independent sources within the network, and place stringent requirements on both the single-mode nature of the sources and the synchronization between them at distant nodes~\cite{kaltenbaek2006interference,kaltenbaek2009swapping,aboussouan2010interference,sun2017swapping}. While the sources themselves must have high spectral purity to ensure indistinguishability across the network~\cite{sun2009indist,changchen2019indist}, photon arrival times must also be synchronized to well within their coherence time, which depending on the physical implementation, may require accuracy to the level of picoseconds~\cite{gerrits2022wrcoexistence}.

In previous work~\cite{lal2024syncsource}, we developed a heralded source of single photons based on spontaneous parametric down-conversion (SPDC), which meets the aforementioned requirements for indistinguishability and can also be synchronized to a network clock. These key capabilities are enabled by the use of an actively mode-locked laser (MLL) as a pump, which provides nearly transform-limited pulses necessary to achieve the highest pair generation rate of highly indistinguishable, single-mode SPDC output photons~\cite{changchen2019indist,ou1997PDCinterference}, and features active stabilization of cavity length to match an external reference clock.

In this work, we characterize the timing stability of two such MLL systems synchronized over optical fiber in a realistic quantum network context. The clock synchronization method employed here is the White Rabbit Precision Time Protocol (WR-PTP), which has been standardized as the high-accuracy precision time protocol (HA-PTP)~\cite{IEEE2020WR_HAPTP,disclaimer}. WR-PTP offer a range of open-source and commercial solutions for picosecond-level synchronization~\cite{lipinski2018WRappl,rizzi2018wrlimits}, and has been deployed in various quantum network prototypes~\cite{alshowkan2022network_wr,alshowkan2022synchronizing,gerrits2022wrcoexistence,rahmouni2024entdist}. Sub-picosecond synchronization of MLLs has been previously demonstrated over several kilometers of optical fiber~\cite{hudson2006synclaser,kim2008syncsources}, but the WR-PTP method explored here provides significant advantages for scalability. In particular, the coexistence of classical WR timing signals with quantum payloads allows for more robust compensation of path length fluctuations that naturally occur over deployed fiber due to environmental noise~\cite{burenkov2023sync,rahmouni2024entdist,mckenzie2023wrsync}. We measure the time deviation (TDEV) for the output pulses of two MLLs synchronized via WR-PTP, characterizing the short- and long-term stability over both 75 km of spooled fiber and 120 km of deployed fiber in a realistic network topology. In each scenario, synchronization within 4 ps is achieved, meeting the temporal indistinguishability requirements set by the SPDC source in Ref.~\cite{lal2024syncsource}.

\section{Experimental setup}

\begin{figure*}[t]
    \centering
    \centerline{\includegraphics[width=1.2\columnwidth,trim={0 60 60 50},clip]{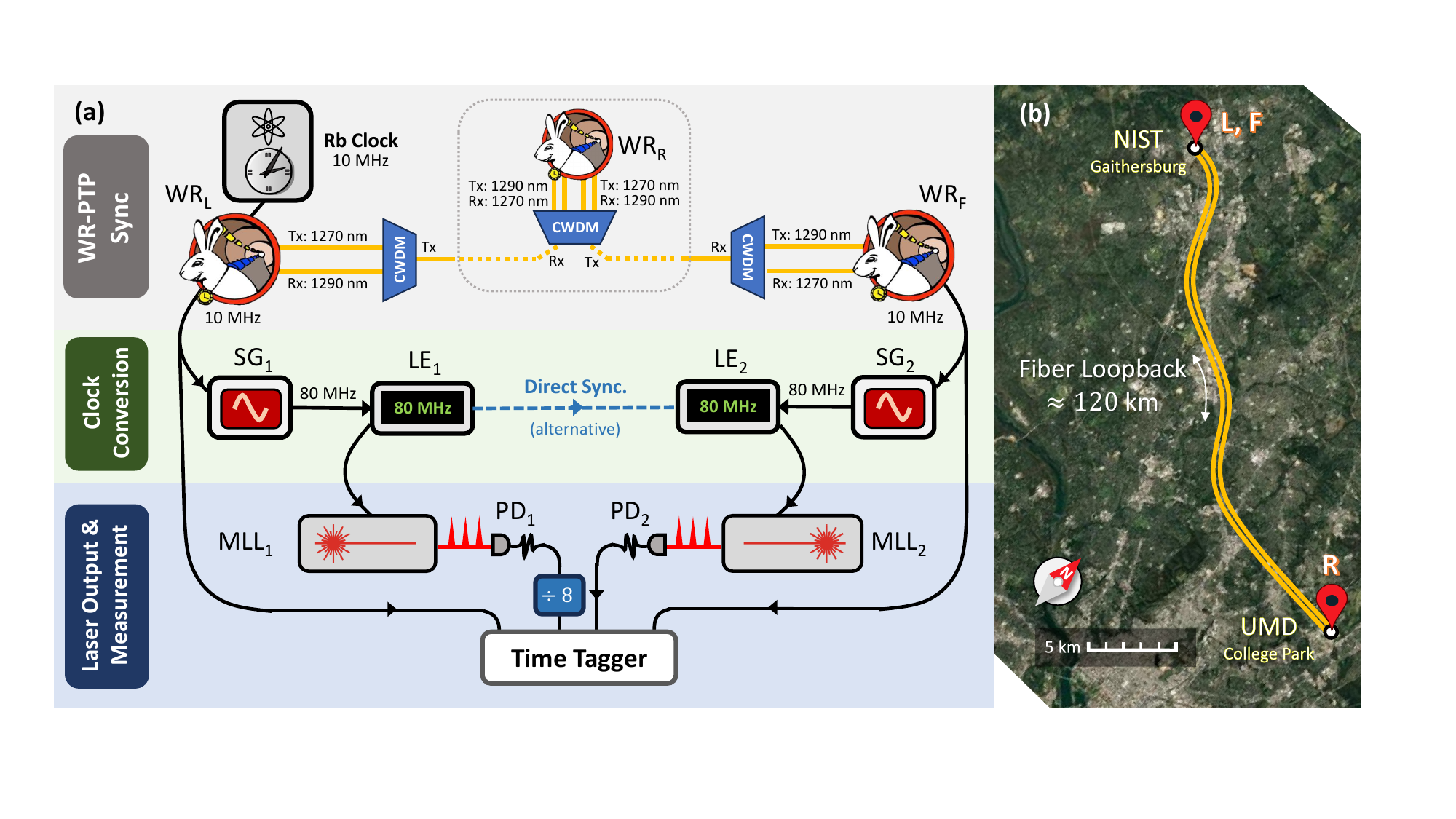}}
    \caption{(a) Schematic of the experimental setup, in three stages. The top layer depicts clock synchronization via the White Rabbit high-precision time transfer protocol. A standard rubidium (Rb) reference clock serves as the network grandmaster clock, and this provides a stable 10 MHz RF signal and 1 pulse-per-second (PPS) signal to a ``leader'' White Rabbit switch $\mathrm{WR_{L}}$.
    In turn, this switch synchronizes a ``follower'' switch $\mathrm{WR_{F}}$ over optical fiber. For a deployed fiber loopback with higher transmission loss (see panel (b)), a ``relay'' node with switch $\mathrm{WR_{R}}$ is synchronized directly to $\mathrm{WR_{L}}$ and serves as an intermediary leader for $\mathrm{WR_{F}}$. On the next layer, the 10 MHz signal from each WR clock is converted into an 80 MHz sinusoid with signal generators $SG_{1,2}$. The 80 MHz signals drive the proprietary locking electronics $\mathrm{LE}_{1,2}$ for each mode-locked laser $\mathrm{MLL_{1,2}}$. The $\mathrm{LE}$ units can also be directly synchronized with local RF signals as an alternative to WR-PTP. The final layer shows the output pulse train of each laser detected by fast photodiodes $\mathrm{PD_{1,2}}$, and a time tagger records these detection signals alongside the original 10 MHz WR clock signals. Time differences between each pair of signals are analyzed in post-processing. WR logo is reproduced with permission, see~\cite{wrcopyright} for copyright and licensing information. (b) Map of the deployed fiber loopback between NIST campus (with leader and follower systems L and F) and UMD, College Park (with relay node R).}
    \label{fig:setup}
\end{figure*}

As a brief overview, this experiment evaluates the relative stability of the pulsed outputs of two MLLs synchronized to the same standard reference clock.
While the lasers are co-located in the same laboratory for direct comparison, the clock signals are distributed over kilometers of optical fiber via the WR-PTP~\cite{lipinski2018WRappl}. Particularly in the case of deployed fiber exposed to environmental noise, this effectively simulates the lasers' performance at spatially separated nodes in a realistic network application~\cite{mckenzie2024dcqnetsync}.
A simplified schematic of the experimental setup is shown in Fig.~\ref{fig:setup}(a). From top to bottom, each layer depicts the distribution of a network clock via WR-PTP, followed by intermediate RF signal conversion to synchronize laser locking electronics to each WR switch, and finally the time-resolved measurement of the optical output pulses of each laser. These steps are described in greater detail below. 

A rubidium (Rb) atomic clock frequency standard serves as our network clock, shown in the top left corner of Fig.~\ref{fig:setup}. This provides a 10 MHz sinusoidal reference signal, connected as a direct input to a low-jitter WR switch, designated here as the ``leader'' $\mathrm{WR_{L}}$. The leader establishes two-way communication with a second low-jitter switch, ``follower'' $\mathrm{WR_{F}}$, over an optical fiber link. The switches interface via small form-factor pluggable (SFP) transceiver modules at wavelengths 1270 nm and 1290 nm, multiplexed over a single fiber with a pair of coarse wavelength division multiplexers (CWDMs), with transmit (Tx) and receive (Rx) ports as indicated in the figure. These wavelengths were chosen with coexistence of classical synchronization signals and quantum signals in mind, where relatively high-intensity classical signals are far detuned from the quantum channel (C-band) to minimize cross-talk~\cite{burenkov2023sync,rahmouni2024entdist}. This fiber connection was tested in two configurations. First, the primary link (dotted yellow line in Fig.~\ref{fig:setup}(a)) consists of interconnected fiber spools up to a length of 75 km.
Second, the primary link consists of deployed fiber in a loopback configuration between NIST and the University of Maryland (UMD), College Park, with a total length of $\approx$120 km as shown in Fig.~\ref{fig:setup}(b).
To accommodate for the additional loss compared to the 75 km spooled fiber, an additional ``relay'' switch $\mathrm{WR_{R}}$ is included at UMD, as shown in the box at the top of Fig.~\ref{fig:setup}(a) and the bottom of the map in panel (b). This switch is synchronized directly to $\mathrm{WR_{L}}$ and serves as an intermediary leader (timeTransmitter~\cite{ieee2023terms}) for $\mathrm{WR_{F}}$. In addition to representing the realistic environmental conditions of a deployed fiber link, this configuration also mimics the topology of a basic repeater node, in which a swapping operation is performed at a central location so that entanglement could be established between the leader and follower nodes~\cite{azuma2023repeaters}.

The leader and follower WR switches each produce TTL pulses at 10 MHz, which are then converted to 80 MHz sinusoidal signals by low-jitter signal generators ($\mathrm{SG_{1,2}}$ in the second layer of Fig.~\ref{fig:setup}) to serve as an external reference for each laser. The lasers used in this experiment are commercial Ti:Sapphire MLL systems, each of which includes proprietary locking electronics (LE) to synchronize to an external clock~\cite{disclaimer}. These MLLs incorporate a phase-locked loop (PLL) control circuit to actively stabilize the cavity length with a piezo-controlled intracavity mirror. When testing the performance of WR-PTP, each LE unit is driven independently by the leader and follower WR switches as shown. However, they can alternatively be synchronized locally by a direct electronic connection, shown by the dotted blue arrow in Fig.~\ref{fig:setup}. As a point of comparison, this ``Direct Sync'' option is also characterized; the leading switch $\mathrm{WR_{L}}$ is used to lock the first unit $\mathrm{LE_{1}}$, and in this case $\mathrm{LE_{2}}$ receives a duplicate reference signal via coaxial cable.

The final layer of Fig.~\ref{fig:setup} shows the measurement and data collection process. Each MLL produces pulses with $\approx$10 ps full width at half maximum (FWHM) and center wavelength 775 nm at a synchronized 80 MHz repetition rate. Each pulse train is detected by a fast silicon photodiode $\mathrm{PD_{1,2}}$ (2 GHz bandwidth), while part of the laser output is also tapped and sent to an autocorrelator (not shown) to verify the pulse width and mode-locking stability of each system. The detector outputs are sent to a Becker \& Hickl SPC-180NX time tagger~\cite{disclaimer}, which has a nominal 1.6 ps root mean square (RMS) jitter and 80 ns dead time. Because the dead time corresponds to a maximum count rate of $\approx$12.5 MHz, one of the signals ($\mathrm{PD_1}$) is divided down to 10 MHz, while the other channel still admits an 80 MHz signal to synchronize the time tagger.
Additionally, the original 10 MHz clock signals from each WR switch are split and used as alternative time tagger inputs, so comparisons can be made between each laser output $\mathrm{PD_{1,2}}$ and each clock signal $\mathrm{WR_{L,F}}$.

In summary, the apparatus in Fig.~\ref{fig:setup} allows for comparison of MLL synchronization stability in three scenarios: (i) WR-PTP synchronization over fiber spools (75 km); (ii) WR-PTP synchronization over deployed fiber (120 km) with a third, intermediate WR switch; (iii) ``Direct sync'' which locks each laser to the same local RF signal. In the next section, the relative time stability between all relevant signals will be characterized in each setting via the time deviation (TDEV) over a range of averaging times~\cite{riley2008handbook}.

\section{Results and discussion}
In the context of network-compatible single-photon and entangled-photon sources, the requirement for temporal indistinguishability between independent sources sets an upper bound on the required clock synchronization jitter (see Fig.~\ref{fig:jitter}). In previous work, the presently considered MLL laser system was used to produce Fourier-transform-limited heralded single-photons with a temporal width of $\approx$35 ps, through parametric down-conversion from a periodically poled potassium titanyl phosphate (PPKTP) crystal and additional narrowband filtering~\cite{lal2024syncsource}. These system parameters set an upper bound for synchronization jitter to ensure sufficient temporal overlap between photon wavepackets, and thus high-visibility interference between independently generated single photons. For example, an RMS jitter of 10 ps between independent single photon sources limits the achievable Hong-Ou-Mandel (HOM) interference visibility to $\approx90$~\% , as shown in Appendix~A.

\begin{figure*}[h]
    \centering
    \begin{subfigure}[h]{0.49\textwidth}
        \centering
        \includegraphics[height=1.9in]{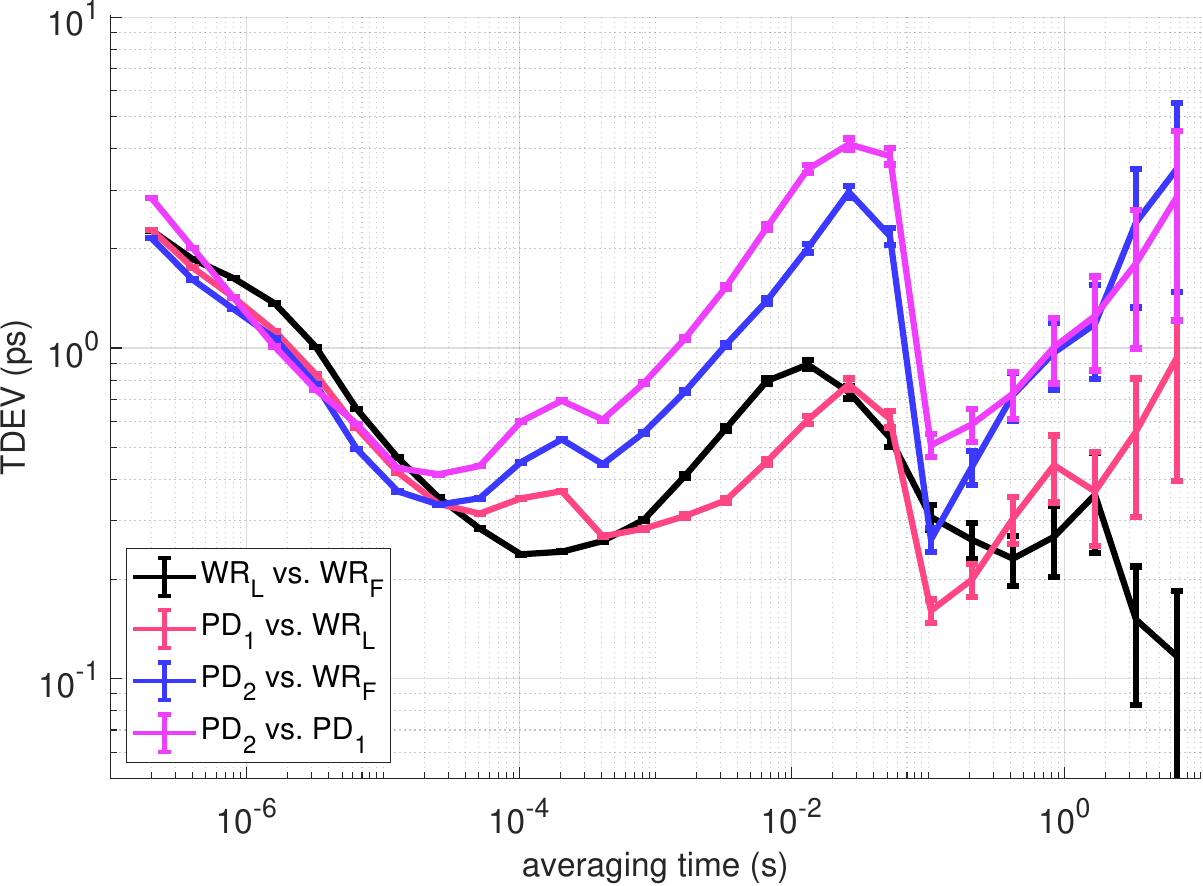}
        \caption{WR-PTP: fiber spool link (75 km)}
    \end{subfigure}
    ~ 
    \begin{subfigure}[h]{0.49\textwidth}
        \centering
        \includegraphics[height=1.9in]{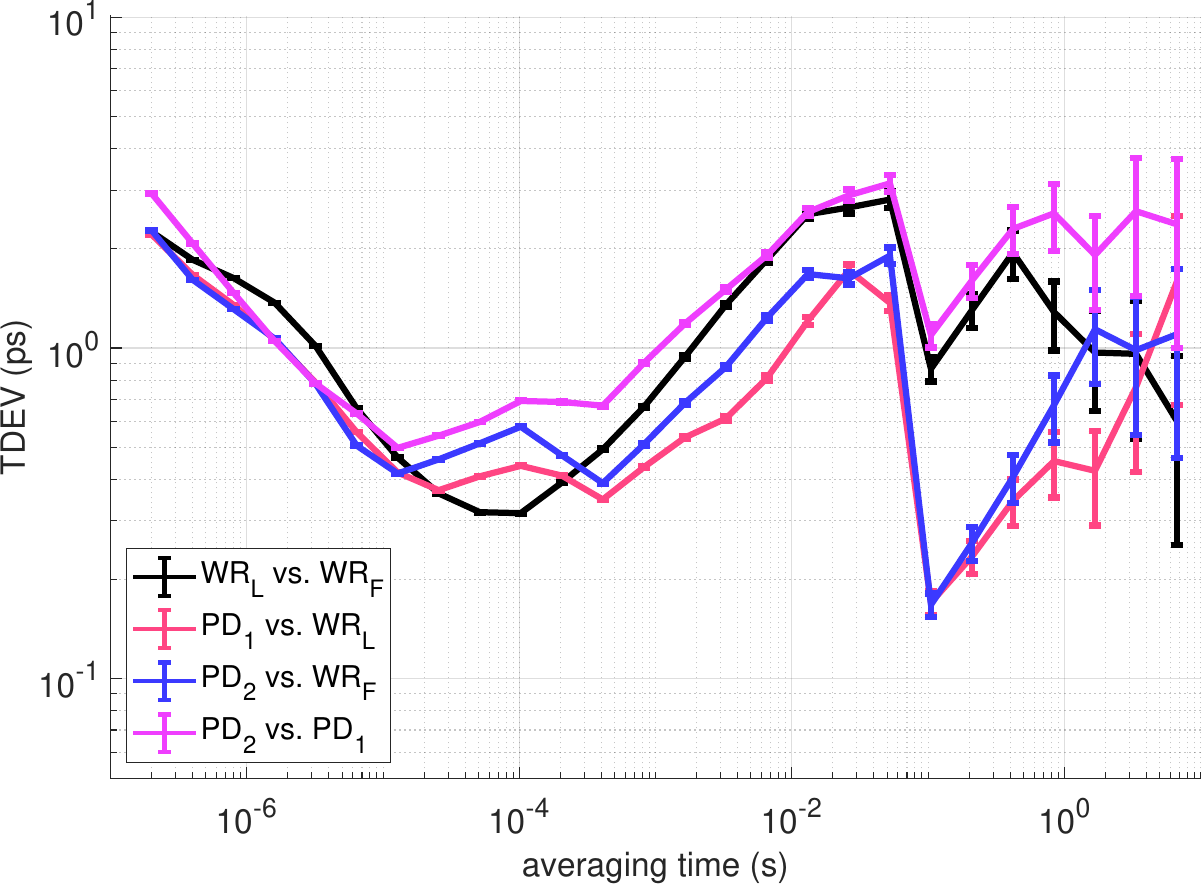}
        \caption{WR-PTP: deployed fiber link (125 km)}
    \end{subfigure}
    ~
    \begin{subfigure}[h]{0.49\textwidth}
        \centering
        \includegraphics[height=1.9in]{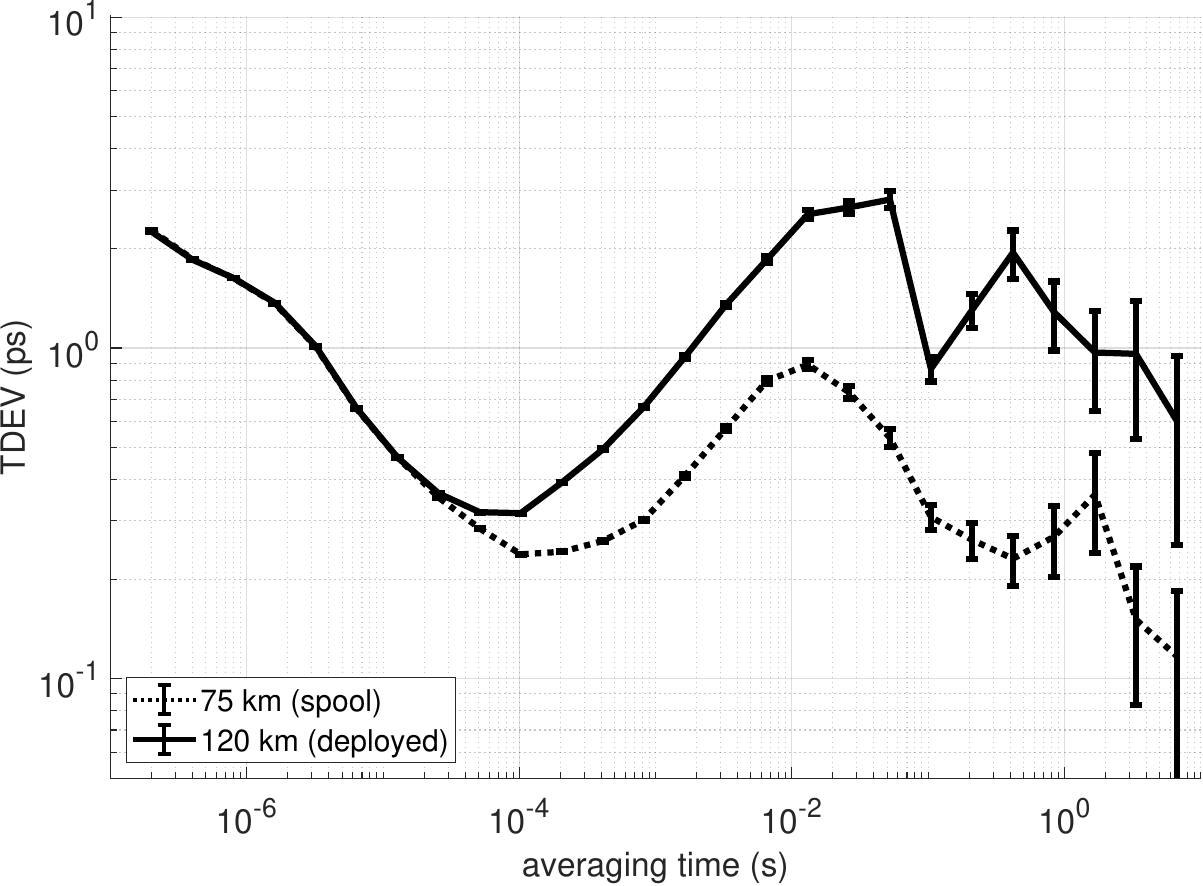}
        \caption{WR Clock-to-Clock comparison: spool vs. deployed}
    \end{subfigure}%
    ~
    \begin{subfigure}[h]{0.49\textwidth}
        \centering
        \includegraphics[height=1.9in]{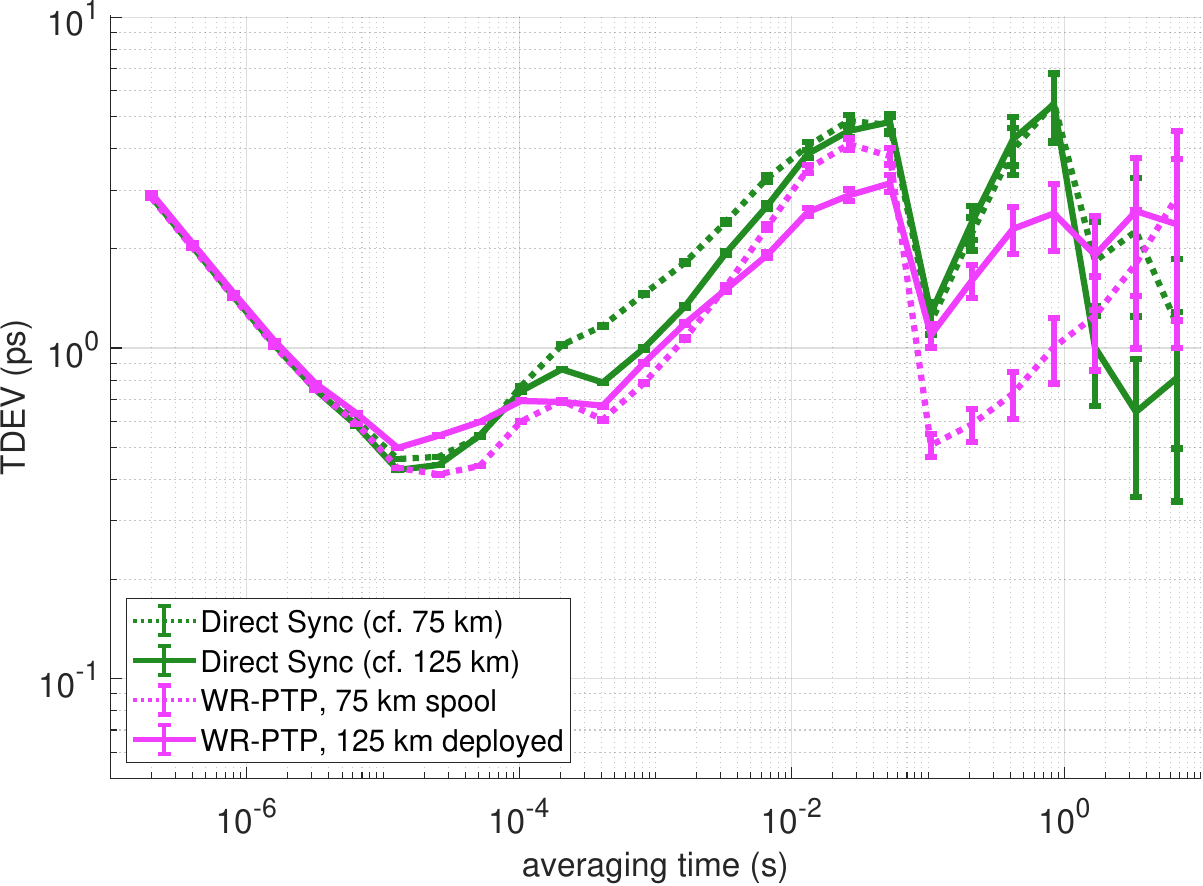}
        \caption{Laser-to-Laser comparison: WR-PTP vs. local sync.}
    \end{subfigure}%
    \caption{TDEV of the clock and laser synchronization using (a) WR-PTP over 75 km spooled fiber, (b) WR-PTP over 120 km deployed fiber, and (d) local synchronization with RF signals; panel (c) directly compares the WR-PTP performance shown in (a) and (b). Error bars represent a 1-$\sigma$ confidence interval for the TDEV estimates. Labels for WR clocks ($\mathrm{WR_{L,F}}$) and fast photodiode signals ($\mathrm{PD_{1,2}}$) match those shown in Fig.~\ref{fig:setup}.}\label{fig:tdev_main}
\end{figure*}

Here, we use the time deviation (TDEV) as a statistical measure of the short- and long-term stability of the laser synchronization~\cite{riley2008handbook}. Measurements are performed for all three configurations described in the previous section: (1) WR-PTP over a fiber spool; (2) WR-PTP over deployed fiber; (3) Direct synchronization with a shared local RF signal. Each set of measurements characterizes pairwise synchronization from clock-to-clock, clock-to-laser, and laser-to-laser, using data from 10 s of accumulated time tags. The main results are summarized in Fig.~\ref{fig:tdev_main} and discussed in further detail below. Additional data relevant to sources of electronic noise, dependencies on fiber length, and further characterization of long-term synchronization stability are included in the Appendix~B.

\subsection{WR-PTP, Fiber Spool Link}

In the first configuration, the leader and follower WR switches are connected by spools of fiber up to 75 km. As long as there was sufficient optical power to initiate WR-PTP ($\gtrsim$0.8 \unit{\micro\watt} for our SFP transceivers), synchronization performance was found to have a minimal dependence on fiber length, so Fig.~\ref{fig:tdev_main}(a) shows data only from the longest tested fiber spool (75 km). Further discussion of the effects of fiber length is included in Appendix~B.

The black curve in Fig~\ref{fig:tdev_main}(a) characterizes the baseline clock-to-clock synchronization between switches $\mathrm{WR_{L}}$ and $\mathrm{WR_{F}}$. The jitter between subsequent 10 MHz clock pulses, given by the leftmost TDEV value, is 2.3 ps. The intrinsic jitter of the time tagger was measured to be 1.7 ps, while the additional 0.6 ps can be largely attributed to the irregular pulse shape of the inverted 10 MHz clock signals (see Appendix~B for the relevant data). At longer averaging times, the clock-to-clock TDEV curve shows a local maximum at millisecond-scale averaging times. This ms-scale noise is a common feature of WR clock recovery systems, likely attributed to phase noise in the SFP Gigabit transceivers and phase detectors~\cite{burenkov2023sync,rizzi2018wrlimits,neelam2024wrphasenosie}. Despite this feature, no longer-term instabilities are observed, and the WR clock signals remain synchronized to within 3 ps deviation for all considered averaging times.

The same plot includes two data sets to characterize clock-to-laser synchronization. The red curve corresponds to signals from the leader system $\mathrm{PD_1}$ vs. $\mathrm{WR_{L}}$, and the blue curve corresponds to the follower system $\mathrm{PD_2}$ vs. $\mathrm{WR_{F}}$. Qualitatively, the latter system shows overall larger TDEV, reaching 3 ps at ms-scale averaging times and similar magnitude drift at longer time scales (1 s to 10 s). Meanwhile, TDEV for the leader system remains sub-picosecond at these averaging times, showing similar overall performance to the clock-to-clock synchronization (black curve), even exhibiting slightly lower ms-scale noise features. As shown in Appendix~B, swapping the roles of ``leader'' and ``follower'' does not significantly change the synchronization performance for each laser system; $\mathrm{MLL_2}$ still exhibits larger TDEV, even when connected to the leader switch $\mathrm{WR_{L}}$. Instead, the noise features observed at $10^{-4}$ s and $10^{-2}$ s averaging times are related to the precise conditions of each laser cavity, subject to nonoptimal alignment and temperature drift, and the (unknown) properties of the proprietary locking electronics. Another TDEV feature of note is the sharp decrease around $10^{-1}$ s averaging times, which is an intrinsic feature of the time tagger (see Appendix~B for relevant data).

Finally, the TDEV for laser-to-laser synchronization is shown in magenta. This measurement displays 2.9 ps jitter, which exceeds that of the other measurements (2.3 ps) due to the inclusion of a pulse counter to divide $\mathrm{PD_1}$ down to a 10 MHz signal. Unsurprisingly, the TDEV between output laser pulses is limited by the underlying laser-to-clock synchronization, and thus lies strictly at or above the largest TDEV values observed in the companion red and blue curves (within the bounds of uncertainty). This results in a peak TDEV value of 4 ps at millisecond-scale averaging times. However, microsecond-scale averaging times show a marginal decrease in TDEV compared to what was measured for the underlying clock signals. The noise from WR switches at this timescale does not appear to transfer to the laser outputs downstream. This suggests the laser PLL circuits do not have the bandwidth to track fast deviations over fewer than $\approx1000$ clock pulses, and hence this noise is filtered out. A more prominent example of this is shown in Appendix~B.

\subsection{WR-PTP, Deployed Fiber Link}

Synchronization for the 120 km deployed fiber link is characterized in Fig.~\ref{fig:tdev_main}(b), with a direct comparison to the 75 km fiber spool measurement shown in plot (c) for the clock-to-clock synchronization ($\mathrm{WR_{L}}$ and $\mathrm{WR_{F}}$). The same characteristic noise features appear at millisecond-scale averaging times in Fig.~\ref{fig:tdev_main}(c), but the peak TDEV value has increased to approximately 3 ps for the deployed fiber link. This difference can be attributed to the introduction of the intermediary switch $\mathrm{WR_{R}}$, which of course introduces the instabilities of a new WR clock discipline algorithm, and also doubles the number of SFP transceivers and their associated phase noise. As seen in Appendix~B, this phase noise also generally increases as the received optical power at each transceiver decreases. The deployed link does have approximately 2 dB lower transmission (one way) compared to the fiber spool, due to additional connector and component losses, but the corresponding effect on TDEV would be relatively insignificant ($<1$ ps).

The full results for the deployed fiber configuration in Fig.~\ref{fig:tdev_main}(b), however, do not show strictly larger TDEV compared to synchronization over the fiber spool. Despite increased noise in the underlying clock-to-clock synchronization, the observed laser-to-laser synchronization ($\mathrm{PD_1}$ vs. $\mathrm{PD_2}$) actually has improved performance compared to that in Fig~\ref{fig:tdev_main}(a), with lower overall TDEV $\lesssim$3 ps. This can be attributed to the greater stability in the follower laser system $\mathrm{MLL_2}$, shown by the blue curve ($\mathrm{PD_2}$ vs. $\mathrm{WR_{F}}$). This measurement was taken on a different day from the previous fiber spool measurement, after which minor adjustments to cavity alignment were made to re-initiate mode locking in both lasers. This evidently improved overall stability in $\mathrm{MLL_2}$ (blue curve), and slightly decreased the stability of $\mathrm{MLL_1}$ (red curve) compared to the previous measurement. As a result, the TDEV for laser-to-clock synchronization for both systems is relatively low ($<$1 ps) for longer averaging times, and the final laser-to-laser synchronization is instead primarily limited by the synchronization between WR clocks ($\mathrm{WR_L}$ vs $\mathrm{WR_F}$). The laser cavities only introduce sub-picosecond deviations around $10^{-4}$ s averaging times.

The observed maximum TDEV of 4 ps lies well within the 35 ps coherence time for the designed single photon sources~\cite{lal2024syncsource}. Following the analysis in Appendix~A, and approximating these short-timescale deviations as 4 ps of RMS timing jitter, this corresponds to an achievable HOM visibility of 98~\%. However, the deployed fiber is subject to temperature changes and other environmental noise that result in additional WR clock drift over longer timescales. The long-term stability of this particular link was previously characterized in Refs.~\cite{mckenzie2023wrsync,mckenzie2024dcqnetsync}, measuring the leader-follower WR delay over the course of several days. Due to the exposed stretches of aerial fiber along the deployed link, peak-to-peak variations of $\approx$200 ps were observed, correlated with changes in temperature. For synchronized quantum network sources, this variable path length must be accounted for. Here, this can be achieved by simply adjusting the local signal generator phase, i.e. with no need for path length compensation, c.f.~\cite{sun2017swapping}. Because the WR-PTP implementation here permits coexistence between the classical synchronization (O-band) and quantum signals (C-band), the classically measured changes in path delay can be directly incorporated into active delay compensation for the quantum signals.

\subsection{Direct Sync, Local RF Link}
In the ``direct sync'' configuration, the 80 MHz reference signal from $\mathrm{SG_{1}}$ to $\mathrm{LE_{1}}$ is also delivered to $\mathrm{LE_{2}}$ by a 5 m copper-shielded coaxial cable, so both lasers are locked to the same local RF signal. The results for laser-to-laser synchronization are shown in Fig.~\ref{fig:tdev_main}(d), alongside the WR-PTP measurements for comparison (dotted lines). Two trials of the direct sync method are included, which were taken immediately before each set of WR-PTP measurements (fiber spool and deployed fiber) and thus account for changes in laser mode-locking stability. In each case, WR-PTP slightly out-performs local synchronization. The most notable change is additional noise at 1 second averaging times, corresponds to slow $<1$ Hz oscillations that can be observed in the free-running histogram during data collection. Of course, this does not represent the limits of local synchronization; both MLLs could be located on the same optical table with shorter RF links, likely improving the relative stability. However, this is less relevant to the objective of simulating synchronized MLLs in different laboratories.

\section{Conclusion}
We have successfully demonstrated the synchronization of two actively mode-locked laser systems over 120 km of deployed optical fiber with the WR-PTP protocol. For a two-switch spooled fiber configuration and three-switch deployed fiber configuration, the relative stability of each laser output was characterized by a TDEV within 4 ps for averaging times up to 10 s. The deployed link topology mirrors that of a practical entanglement swapping scheme, and the link itself includes aerial fiber subject to environmental noise as in a realistic network application. For the SPDC sources with 35 ps coherence time considered here, the low timing jitter between each pump laser enables high temporal indistinguishability between photon wavepackets, and thus high-visibility interference between independent sources. The synchronization signals considered here are also capable of coexisting with quantum signals in the C-band along the same optical fiber. In principle, this allows long-term drift between network clocks to be corrected with path delay compensation. The synchronization capability demonstrated here further enhances the utility of the high-quality sources of indistinguishable photons introduced in Ref.~\cite{lal2024syncsource}, enabling essential communication protocols such as entanglement swapping in a scalable manner, suitable for the increasing complexity of metropolitan-scale quantum networks.

\appendix

\section*{Appendix A: Temporal Indistinguishability}

\begin{figure*}[h]
    \centering
    \begin{subfigure}[h]{0.51\textwidth}
        \centering
        \includegraphics[height=1.7in,trim={0 0 0 0},clip]{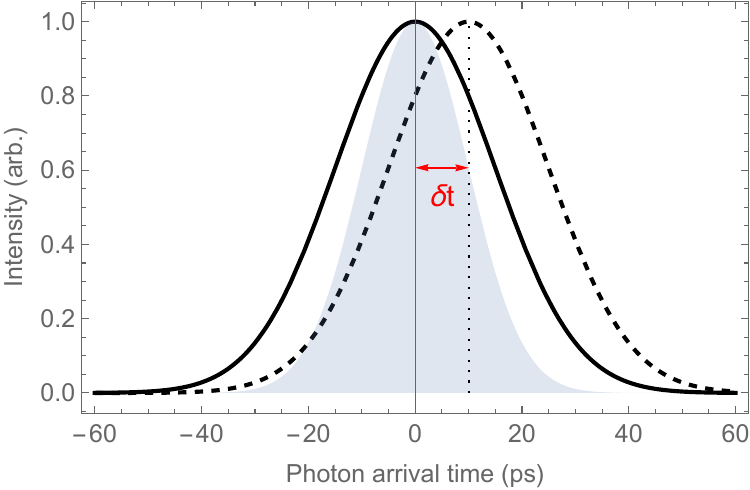}
        \caption{Timing jitter of single-photon wavepackets}
    \end{subfigure}%
    ~
    \begin{subfigure}[h]{0.51\textwidth}
        \centering
        \includegraphics[height=1.75in,trim={0 0 0 15},clip]{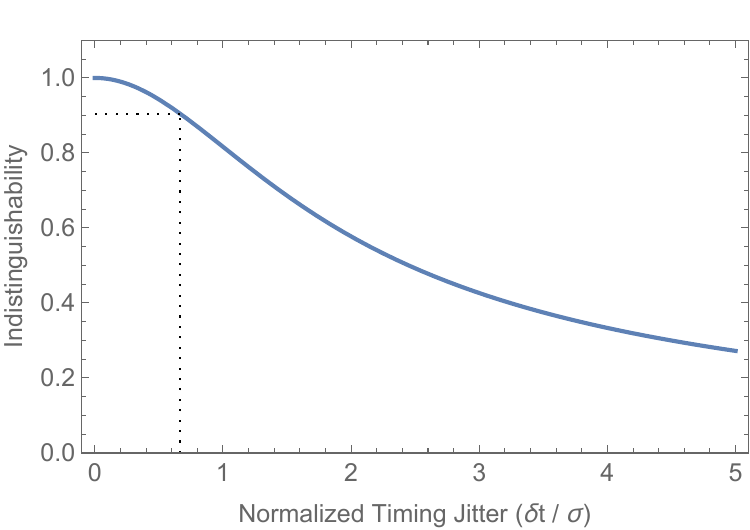}
        \caption{Effect of timing jitter on indistinguishability}
    \end{subfigure}
    \caption{(a) Visual representation of single-photon wavepackets with coherence time of 35 ps (FWHM) subject to RMS timing jitter $\delta t=10$ ps. The shaded envelope shows a Gaussian distribution of relative delays between photon arrival times, while the solid and dotted curves illustrate temporal overlap of wavepackets offset by $\delta t$ delay. (b) Single-photon indistinguishability as a function of timing jitter $\delta t$, normalized by wavepacket RMS width $\sigma$. The dotted line corresponds to the values in panel (a), giving a limit of $\approx90$~\% HOM interference visibility for these paramaters.}\label{fig:jitter}
\end{figure*}

This appendix quantifies the effects of timing jitter in photon arrival times on the temporal indistinguishability of single photon sources. In the context of Hong-Ou-Mandel (HOM) interference between photons from independent, synchronized sources, it can be shown that the measured indistinguishability has the following dependence on RMS timing jitter $\delta t$~\cite{burenkov2023sync}:
\begin{equation}\label{eq:indist}
    I = \left(1+\frac{\delta t^2}{\sigma^2}\right)^{-1/2}
\end{equation}
where $\sigma$ is the RMS width of the gaussian photon wavepackets. For the particular SPDC source considered here, heralded single photons have a measured coherence time of 35~ps (FWHM), or $\sigma=15$~ps. Fig.~\ref{fig:jitter}(a) illustrates the temporal overlap of these wavepackets subject to $\delta t=10$~ps of timing jitter. The effect on indistinguishability from Eq.~\ref{eq:indist} is shown in Fig.~\ref{fig:jitter}(b) for a range of parameters. The which-path information introduced by imperfect temporal overlap limits the achievable HOM visibility-- e.g., $\approx90~\%$ for $\delta t=10$~ps of timing jitter.

\section*{Appendix B: WR-PTP Noise Sources}
\begin{figure*}[h]
    \centering
    \begin{subfigure}[h]{0.49\textwidth}
        \centering
        \includegraphics[height=1.9in]{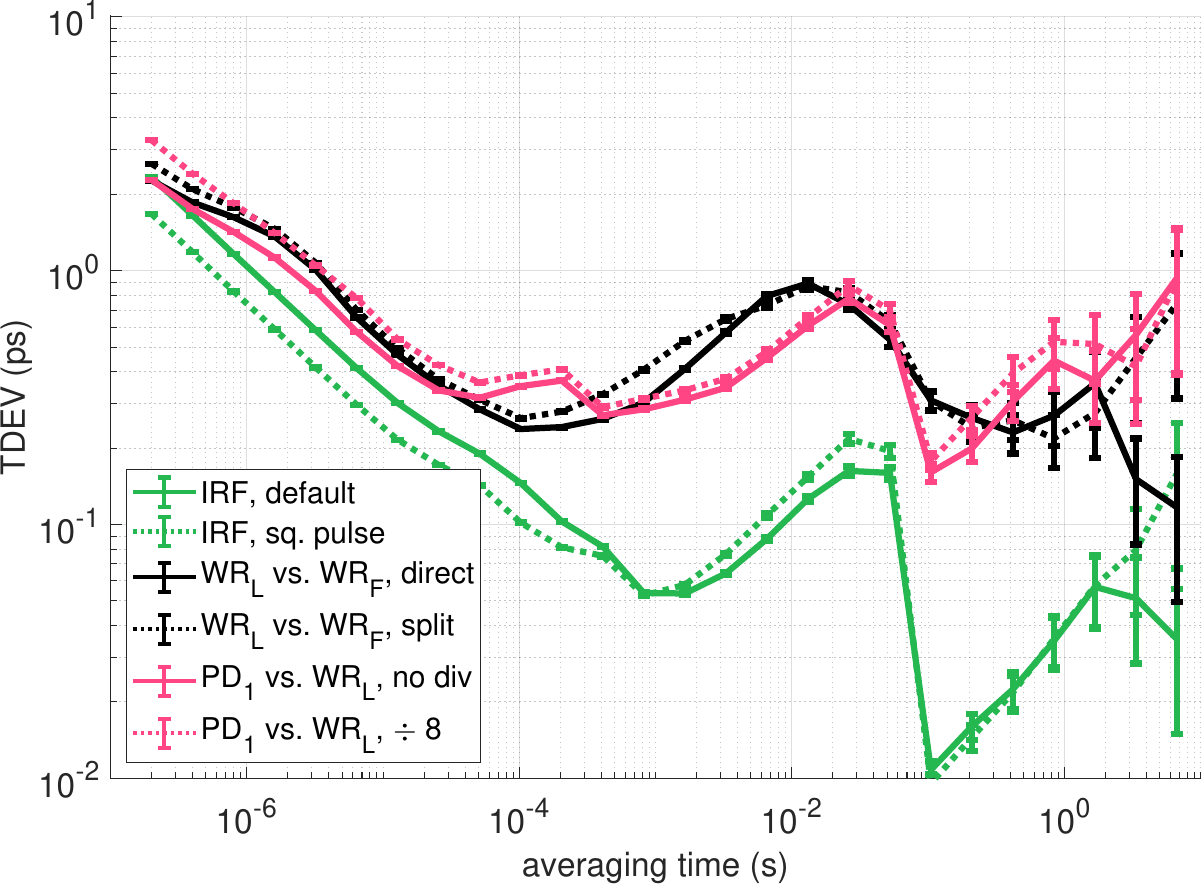}
        \caption{Effect of electronic components (IRF; 75 km spool).}
    \end{subfigure}%
        ~ 
    \begin{subfigure}[h]{0.49\textwidth}
        \centering
        \includegraphics[height=1.9in]{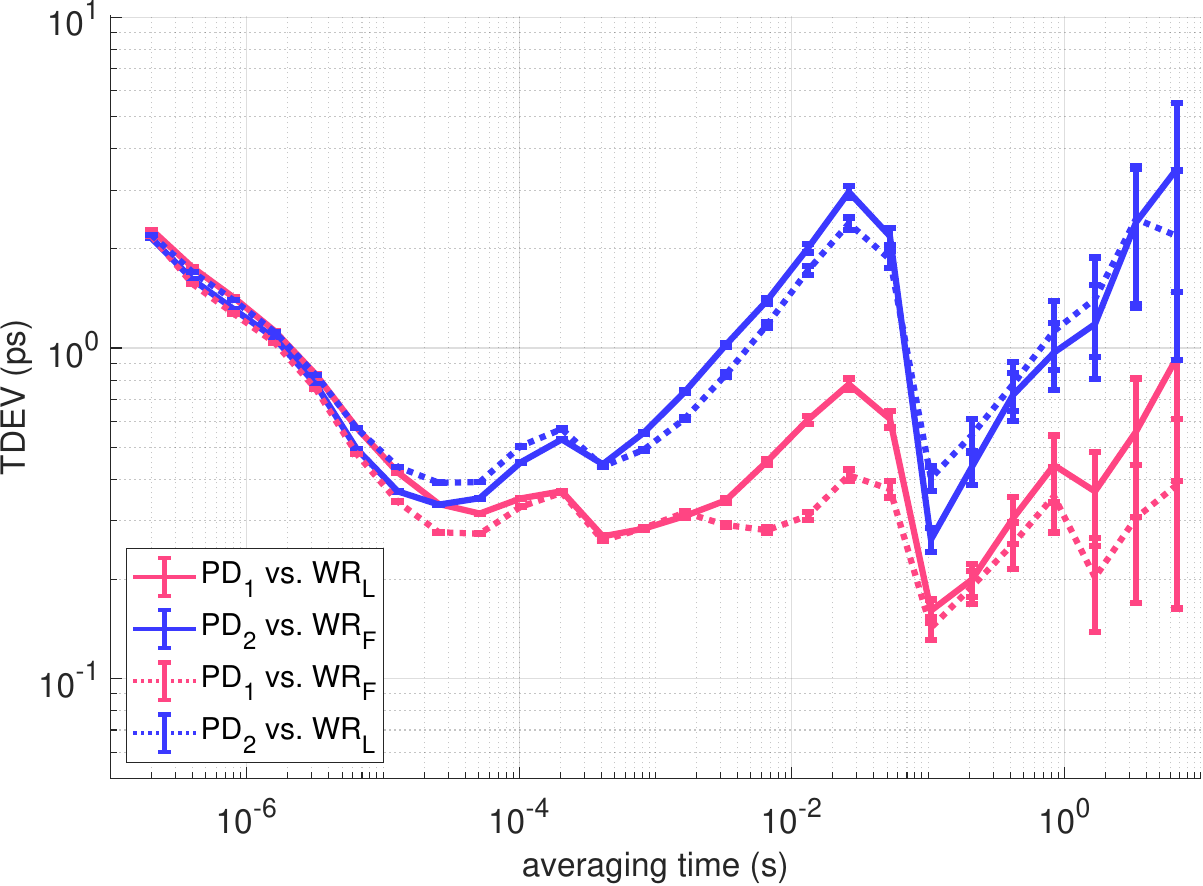}
        \caption{Effect of leader-follower role swap (75 km spool).}
    \end{subfigure}
        ~ 
    \begin{subfigure}[h]{0.49\textwidth}
        \centering
        \includegraphics[height=1.9in]{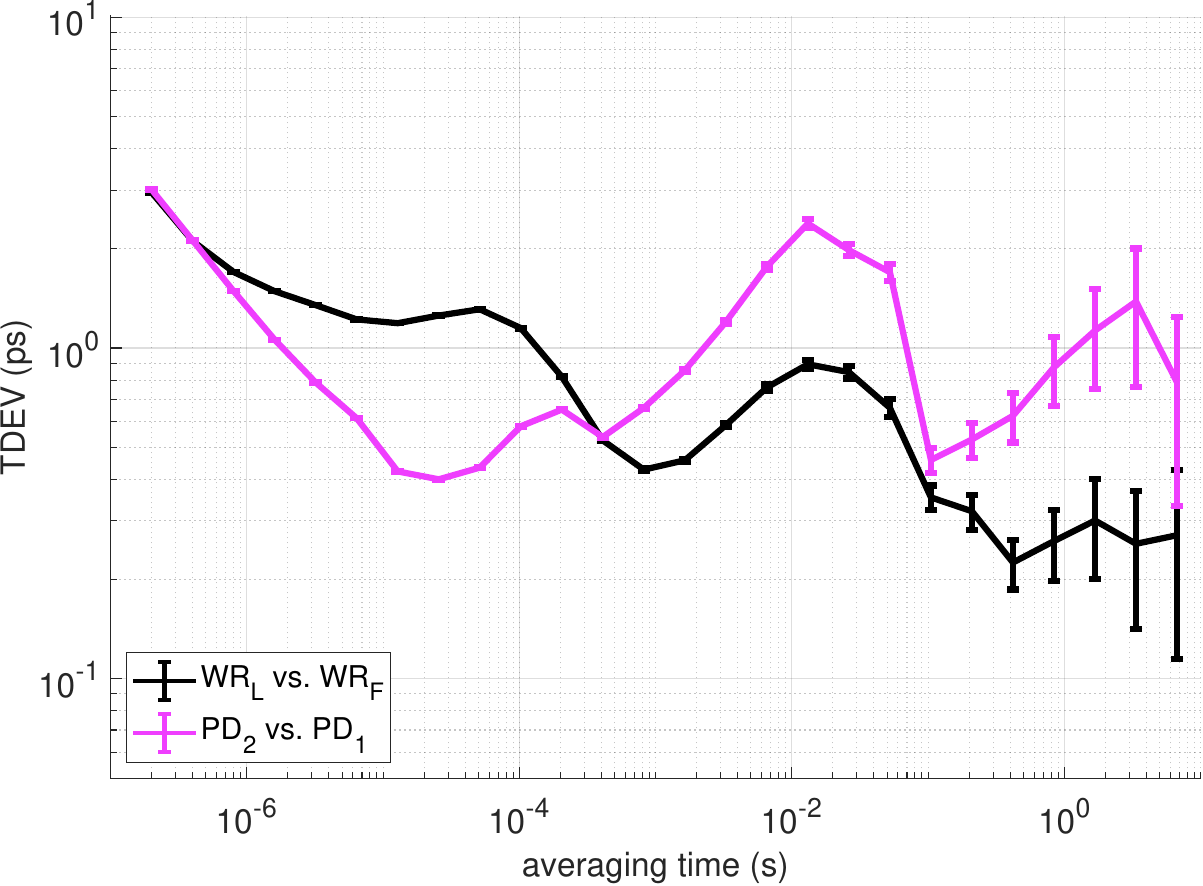}
        \caption{Effect of high-frequency clock noise (50 km spool)}
    \end{subfigure}
        ~ 
    \begin{subfigure}[h]{0.49\textwidth}
        \centering
        \includegraphics[height=1.9in]{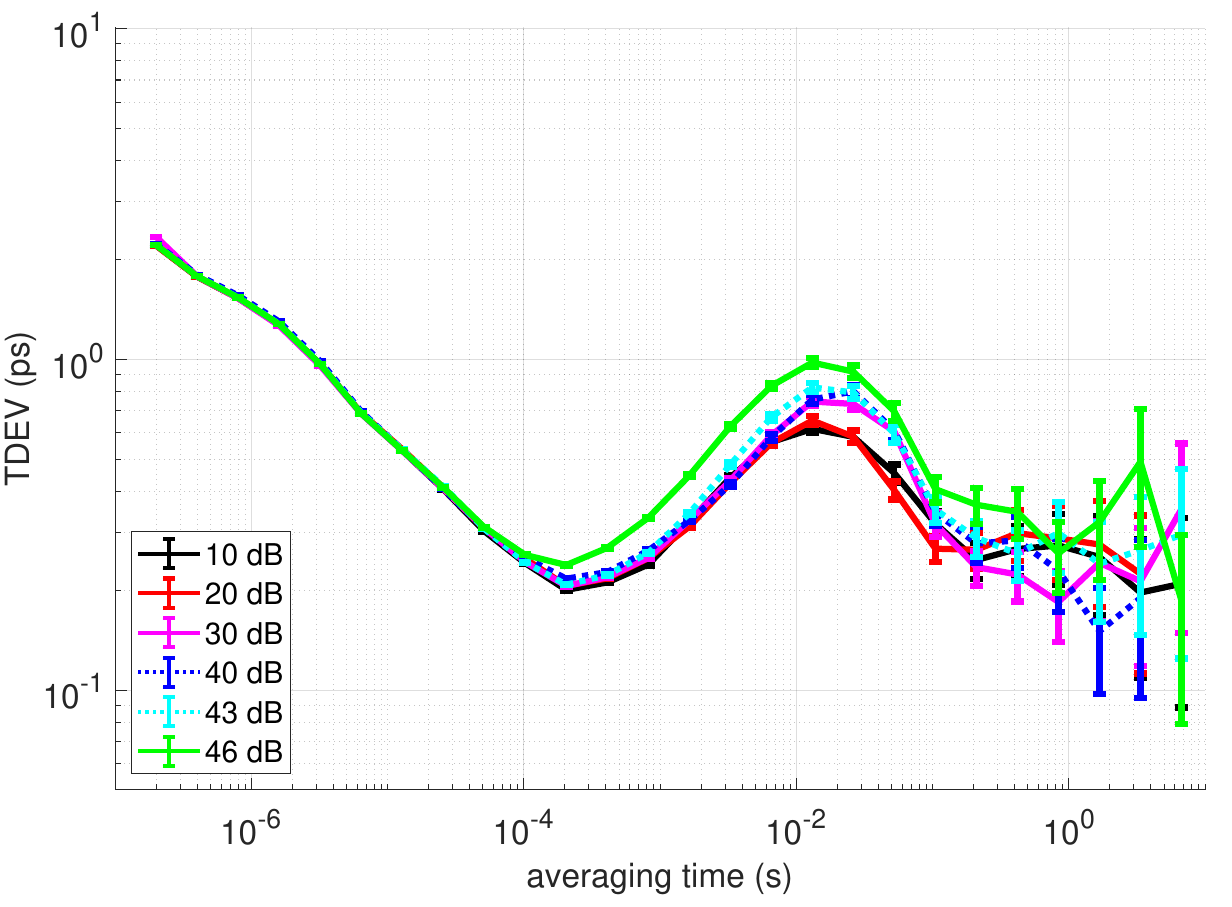}
        \caption{Effect of SFP transceiver attenuation (clock-to-clock).}
    \end{subfigure}%
    \caption{WR-PTP synchronization performance-- comparison for various configurations. Error bars represent a 1-$\sigma$ confidence interval for the TDEV estimates. Details in Appendix~B.}\label{fig:tdev_extra}
\end{figure*}

In this appendix, we provide further details on the sources of noise in the WR-PTP configurations considered in this study. The relevant TDEV plots are shown in Figure~\ref{fig:tdev_extra}.

In panel (a), the instrument response function (IRF) for the time tagger is shown, along with several data sets meant to summarize the sources of electronic jitter. First, the IRF was measured by splitting the TTL pulses of one WR clock output, sent into both time tagger channels. The green dotted line shows the result when these square pulses are measured directly, resulting in 1.7 ps jitter (leftmost TDEV value), close to the nominal RMS value of 1.6 ps. However, comparing different clock signals requires the use of additional RF components (inverter, DC blocker, and attenuator), present in all main clock signal measurements. The ``default'' IRF measurement (solid green) includes these components, showing a baseline jitter of 2.3 ps. In both IRF measurements, a sharp drop occurs at $10^{-1}$ s averaging times, and this feature also appears in the majority of data sets.

The remaining data in panel (a) corresponds to WR-PTP synchronization over the 75 km fiber spool. The solid black curve corresponds to when the two WR clock signals are directly input to the time tagger and not connected to either laser system, resulting in 2.3 ps jitter just like the IRF. The dotted black curve shows the case when both clock signals are split and routed to each laser (see Fig.~\ref{fig:setup}), as is the case for all laser-to-clock synchronization measurements. This slightly increases jitter to 2.6 ps. Finally, the effects of the pulse counter (frequency divider) are shown by two measurements of $\mathrm{PD_1}$ vs $\mathrm{WR_L}$, with and without this component. Here, the jitter is shown to increase from 2.3 ps to 3.3 ps. Note that the final laser-to-laser synchronization has slightly lower measured jitter (2.9 ps), since the sharp pulses produced by the fast photodiodes $\mathrm{PD_{1,2}}$ can be measured with higher precision than the inverted clock pulses. Importantly, no significant changes are introduced at longer averaging times by any of these components.

In panel (b), laser-to-clock synchronization is shown for WR-PTP over the fiber spool in two configurations: the default shown in Fig.~\ref{fig:setup}, with the leader $\mathrm{MLL_1}$ and follower $\mathrm{MLL_2}$; and the case where the leader and follower roles are swapped. The swapped configuration shows very similar behavior, with $\mathrm{MLL_2}$ still showing the most instability of the two systems despite being connected to the leader WR switch. This suggests the role of the WR switch has a negligible effect on laser synchronization performance, compared to the mode-locking conditions described in the main text.

In panel (c), WR-PTP synchronization over a 50 km fiber spool is characterized using a different set of WR switches. This particular leader switch generated additional noise that increased TDEV for microsecond-scale averaging times. However, this only appeared in measurements of the $\mathrm{WR_L}$ clock signal, as in the shown clock-to-clock synchronization (black), and did not transfer to the laser output synchronization $\mathrm{PD_1}$ vs. $\mathrm{PD_2}$ (magenta). Similar low-pass filtering is also present in the main results of Fig.~\ref{fig:tdev_main}.

Finally, the sequence of measurements in panel (d) shows WR-PTP over a fiber patch cable with various degrees of attenuation. As the attenuation increases and power drops to the minimum level to maintain a lock, the typical millisecond-scale TDEV feature also grows from sub-picosecond level to approximately 1 ps. This particular measurement was made using more sensitive SFP modules, putting the attenuation threshold at 46 dB. The difference between received power in the deployed link compared to the 75 km fiber spool link is relatively small (2 dB), and hence any significant difference between the two is not the result of low optical power, but rather the increased complexity introduced by a third WR switch $\mathrm{WR_R}$.

\begin{backmatter}
\bmsection{Acknowledgments} The authors thank Alessandro Restelli, Joshua Bienfang and Jabir Marakkarakath Vadakkepurayil for the valuable technical guidance and review of this manuscript.

\bmsection{Disclosures}
The authors declare no conflicts of interest.

\bmsection{Data availability}
Data underlying the results presented in this paper are not publicly available at this time but may be obtained from the authors upon reasonable request.

\end{backmatter}


\bibliography{refs_anouar,refs_ivan,refs_nijil,refs_nunn}

\begin{thebibliography}{10}
\newcommand{\enquote}[1]{``#1''}

\bibitem{bienfang2004qkd_sync}
J.~C. Bienfang, A.~J. Gross, A.~Mink, \emph{et~al.}, \enquote{Quantum key distribution with 1.25 gbps clock synchronization,} {\protect\JournalTitle{Opt. Express}} \textbf{12}, 2011--2016 (2004).

\bibitem{cirac1997entdist}
J.~I. Cirac, P.~Zoller, H.~J. Kimble, and H.~Mabuchi, \enquote{Quantum state transfer and entanglement distribution among distant nodes in a quantum network,} {\protect\JournalTitle{Phys. Rev. Lett.}} \textbf{78}, 3221--3224 (1997).

\bibitem{kim2022timebin}
J.-H. Kim, J.-W. Chae, Y.-C. Jeong, and Y.-H. Kim, \enquote{Quantum communication with time-bin entanglement over a wavelength-multiplexed fiber network,} {\protect\JournalTitle{APL Photonics}} \textbf{7}, 016106 (2022).

\bibitem{jin2015swapping}
R.-B. Jin, M.~Takeoka, U.~Takagi, \emph{et~al.}, \enquote{Highly efficient entanglement swapping and teleportation at telecom wavelength,} {\protect\JournalTitle{Scientific Reports}} \textbf{5}, 9333 (2015).

\bibitem{sun2017swapping}
Q.-C. Sun, Y.-F. Jiang, Y.-L. Mao, \emph{et~al.}, \enquote{Entanglement swapping over 100\&\#x2009;\&\#x2009;km optical fiber with independent entangled photon-pair sources,} {\protect\JournalTitle{Optica}} \textbf{4}, 1214--1218 (2017).

\bibitem{pant2019swapping}
M.~Pant, H.~Krovi, D.~Towsley, \emph{et~al.}, \enquote{Routing entanglement in the quantum internet,} {\protect\JournalTitle{npj Quantum Information}} \textbf{5}, 25 (2019).

\bibitem{wang2022entdist}
Y.~Wang, Z.-Y. Hao, Z.-H. Liu, \emph{et~al.}, \enquote{Remote entanglement distribution in a quantum network via multinode indistinguishability of photons,} {\protect\JournalTitle{Phys. Rev. A}} \textbf{106}, 032609 (2022).

\bibitem{kaltenbaek2006interference}
R.~Kaltenbaek, B.~Blauensteiner, M.~Żukowski, \emph{et~al.}, \enquote{Experimental {Interference} of {Independent} {Photons},} {\protect\JournalTitle{Phys. Rev. Lett.}} \textbf{96}, 240502 (2006). Publisher: American Physical Society.

\bibitem{kaltenbaek2009swapping}
R.~Kaltenbaek, R.~Prevedel, M.~Aspelmeyer, and A.~Zeilinger, \enquote{High-fidelity entanglement swapping with fully independent sources,} {\protect\JournalTitle{Phys. Rev. A}} \textbf{79}, 040302 (2009). Publisher: American Physical Society.

\bibitem{aboussouan2010interference}
P.~Aboussouan, O.~Alibart, D.~B. Ostrowsky, \emph{et~al.}, \enquote{High-visibility two-photon interference at a telecom wavelength using picosecond-regime separated sources,} {\protect\JournalTitle{Phys. Rev. A}} \textbf{81}, 021801 (2010).

\bibitem{sun2009indist}
F.~W. Sun and C.~W. Wong, \enquote{Indistinguishability of independent single photons,} {\protect\JournalTitle{Phys. Rev. A}} \textbf{79}, 013824 (2009).

\bibitem{changchen2019indist}
C.~Chen, J.~E. Heyes, K.-H. Hong, \emph{et~al.}, \enquote{Indistinguishable single-mode photons from spectrally engineered biphotons,} {\protect\JournalTitle{Opt. Express}} \textbf{27}, 11626--11634 (2019).

\bibitem{gerrits2022wrcoexistence}
T.~Gerrits, I.~A. Burenkov, Y.~S. Li-Baboud, \emph{et~al.}, \enquote{White {Rabbit}-assisted quantum network node synchronization with quantum channel coexistence,} in \emph{Conference on {Lasers} and {Electro}-{Optics} (2022), paper {FM1C}.2,}  (Optica Publishing Group, 2022), p. FM1C.2.

\bibitem{lal2024syncsource}
N.~Lal, I.~A. Burenkov, Y.-S. Li-Baboud, \emph{et~al.}, \enquote{Synchronized source of indistinguishable photons for quantum networks,} {\protect\JournalTitle{Opt. Express, OE}} \textbf{32}, 18257--18267 (2024). Publisher: Optica Publishing Group.

\bibitem{ou1997PDCinterference}
Z.~Y. Ou, \enquote{Parametric down-conversion with coherent pulse pumping and quantum interference between independent fields,} {\protect\JournalTitle{Quantum and Semiclassical Optics: Journal of the European Optical Society Part B}} \textbf{9}, 599 (1997).

\bibitem{IEEE2020WR_HAPTP}
\enquote{{IEEE} standard for a precision clock synchronization protocol for networked measurement and control systems,} {\protect\JournalTitle{IEEE Std 1588-2019 (Revision of IEEE Std 1588-2008)}} pp. 1--499 (2020).

\bibitem{disclaimer}
Commercial equipment and software referred to in this work is identified for informational purposes only, and does not imply recommendation of or endorsement by the National Institute of Standards and Technology, nor does it imply that the products so identified are necessarily the best available for the purpose.

\bibitem{lipinski2018WRappl}
M.~Lipiński, E.~van~der Bij, J.~Serrano, \emph{et~al.}, \enquote{White rabbit applications and enhancements,} in \emph{2018 IEEE International Symposium on Precision Clock Synchronization for Measurement, Control, and Communication (ISPCS),}  (2018), pp. 1--7.

\bibitem{rizzi2018wrlimits}
M.~Rizzi, M.~Lipinski, P.~Ferrari, \emph{et~al.}, \enquote{White rabbit clock synchronization: Ultimate limits on close-in phase noise and short-term stability due to fpga implementation,} {\protect\JournalTitle{IEEE Transactions on Ultrasonics, Ferroelectrics, and Frequency Control}} \textbf{65}, 1726--1737 (2018).

\bibitem{alshowkan2022network_wr}
M.~Alshowkan, P.~G. Evans, B.~P. Williams, \emph{et~al.}, \enquote{Advanced architectures for high-performance quantum networking,} {\protect\JournalTitle{J. Opt. Commun. Netw.}} \textbf{14}, 493--499 (2022).

\bibitem{alshowkan2022synchronizing}
M.~Alshowkan, P.~G. Evans, B.~P. Williams, \emph{et~al.}, \enquote{Synchronizing a quantum local area network with white rabbit,} in \emph{2022 Conference on Lasers and Electro-Optics (CLEO),}  (2022), pp. 1--2.

\bibitem{rahmouni2024entdist}
A.~Rahmouni, P.~S. Kuo, Y.~S. Li-Baboud, \emph{et~al.}, \enquote{100-km entanglement distribution with coexisting quantum and classical signals in a single fiber,} {\protect\JournalTitle{J. Opt. Commun. Netw., JOCN}} \textbf{16}, 781--787 (2024). Publisher: Optica Publishing Group.

\bibitem{hudson2006synclaser}
D.~D. Hudson, S.~M. Foreman, S.~T. Cundiff, and J.~Ye, \enquote{Synchronization of mode-locked femtosecond lasers through a fiber link,} {\protect\JournalTitle{Opt. Lett.}} \textbf{31}, 1951--1953 (2006). Publisher: Optica Publishing Group.

\bibitem{kim2008syncsources}
J.~Kim, J.~A. Cox, J.~Chen, and F.~X. Kärtner, \enquote{Drift-free femtosecond timing synchronization of remote optical and microwave sources,} {\protect\JournalTitle{Nature Photon}} \textbf{2}, 733--736 (2008). Publisher: Nature Publishing Group.

\bibitem{burenkov2023sync}
I.~A. Burenkov, A.~Semionov, Hala, \emph{et~al.}, \enquote{Synchronization and coexistence in quantum networks,} {\protect\JournalTitle{Opt. Express}} \textbf{31}, 11431--11446 (2023).

\bibitem{mckenzie2023wrsync}
W.~McKenzie, Y.~S. Li-Baboud, M.~Morris, \emph{et~al.}, \enquote{Sub-200 ps {Quantum} {Network} {Node} {Synchronization} over a 128 km {Link} {White} {Rabbit} {Architecture},} in \emph{{CLEO} 2023 (2023), paper {FF3A}.3,}  (Optica Publishing Group, 2023), p. FF3A.3.

\bibitem{wrcopyright}
The White Rabbit logo is the intellectual property of CERN. The logo is licensed under “Attribution-ShareAlike 4.0 International (CC BY-SA 4.0), https://creativecommons.org/licenses/by-sa/4.0/. The logo is authored by Alexandra Lewis.

\bibitem{mckenzie2024dcqnetsync}
W.~McKenzie, A.~M. Richards, S.~Patel, \emph{et~al.}, \enquote{{Clock synchronization characterization of the Washington DC metropolitan quantum network (DC-QNet)},} {\protect\JournalTitle{Applied Physics Letters}} \textbf{125}, 164004 (2024).

\bibitem{ieee2023terms}
\enquote{Ieee standard for a precision clock synchronization protocol for networked measurement and control systems amendment 2: Master-slave optional alternative terminology,} {\protect\JournalTitle{IEEE Std 1588g-2022 (Amendment to IEEE Std 1588-2019 as amended by IEEE Std 1588b-2022)}} pp. 1--14 (2023).

\bibitem{azuma2023repeaters}
K.~Azuma, S.~E. Economou, D.~Elkouss, \emph{et~al.}, \enquote{Quantum repeaters: From quantum networks to the quantum internet,} {\protect\JournalTitle{Rev. Mod. Phys.}} \textbf{95}, 045006 (2023).

\bibitem{riley2008handbook}
W.~Riley and D.~Howe, \enquote{Handbook of frequency stability analysis,}  (2008).

\bibitem{neelam2024wrphasenosie}
Neelam, S.~Kandeepan, and S.~Panja, \enquote{Phase {Noise} {Analysis} of {Time} {Transfer} over {White} {Rabbit}-{Network} {Based} {Optical} {Fibre} {Links},} {\protect\JournalTitle{Sensors}} \textbf{24}, 381 (2024). Number: 2 Publisher: Multidisciplinary Digital Publishing Institute.

\end{thebibliography}

\end{document}